\documentclass[sigconf]{aamas} 

\usepackage{balance} 

\usepackage[acronym]{glossaries}
\usepackage[ruled,vlined,linesnumbered,resetcount]{algorithm2e}
\usepackage{etoolbox}
\DeclareMathOperator*{\argmax}{arg\,max}

\setcopyright{ifaamas}
\acmConference[AAMAS '21]{Proc.\@ of the 20th International Conference on Autonomous Agents and Multiagent Systems (AAMAS 2021)}{May 3--7, 2021}{London, UK}{U.~Endriss, A.~Now\'{e}, F.~Dignum, A.~Lomuscio (eds.)}
\copyrightyear{2021}
\acmYear{2021}
\acmDOI{}
\acmPrice{}
\acmISBN{}

\acmSubmissionID{36}

\title[Exploring the Impact of Tunable Agents\\ in Sequential Social Dilemmas]{Exploring the Impact of Tunable Agents\\ in Sequential Social Dilemmas}
\titlenote{This paper extends our AAMAS 2021 extended abstract \cite{ocallaghan2021tunable} with additional experimental results and analysis.}

\author{David O'Callaghan}
\affiliation{
  \department{School of Computer Science}
  \institution{National University of Ireland Galway}}
\email{ocallaghan1@gmail.com}

\author{Patrick Mannion}
\affiliation{
  \department{School of Computer Science}
  \institution{National University of Ireland Galway}}
\email{patrick.mannion@nuigalway.ie}

\begin{abstract}
When developing reinforcement learning agents, the standard approach is to train an agent to converge to a fixed policy that is as close to optimal as possible for a single fixed reward function. If different agent behaviour is required in the future, an agent trained in this way must normally be either fully or partially retrained, wasting valuable time and resources. In this study, we leverage multi-objective reinforcement learning to create tunable agents, i.e. agents that can adopt a range of different behaviours according to the designer's preferences, without the need for retraining. We apply this technique to sequential social dilemmas, settings where there is inherent tension between individual and collective rationality. Learning a single fixed policy in such settings leaves one at a significant disadvantage if the opponents' strategies change after learning is complete. In our work, we demonstrate empirically that the tunable agents framework allows easy adaption between cooperative and competitive behaviours in sequential social dilemmas without the need for retraining, allowing a single trained agent model to be adjusted to cater for a wide range of behaviours and opponent strategies.
\end{abstract}


\newcommand{\BibTeX}{\rm B\kern-.05em{\sc i\kern-.025em b}\kern-.08em\TeX}

\newacronym{rl}{RL}{reinforcement learning}

\newacronym{morl}{MORL}{multi-objective reinforcement learning}

\newacronym{mas}{MAS}{multi-agent system}

\newacronym{marl}{MARL}{Multi-agent reinforcement learning}

\newacronym{sg}{SG}{stochastic game}

\newacronym[firstplural=Markov decision processes (MDPs)]{mdp}{MDP}{Markov decision process}

\newacronym{momdp}{MOMDP}{multi-objective Markov decision process}

\newacronym{mosg}{MOSG}{multi-objective stochastic game}

\newacronym{ai}{AI}{artificial intelligence}

\newacronym{dqn}{DQN}{deep Q-network}

\newacronym{dst}{DST}{Deep Sea Treasure}

\newacronym{nn}{NN}{neural network}

\newacronym{dnn}{DNN}{deep neural network}

\newacronym{cnn}{CNN}{convolutional neural network}

\newacronym{relu}{ReLU}{rectified linear unit}

\newacronym{mlp}{MLP}{multi-layer perceptron}

\newacronym[firstplural=sequential social dilemmas (SSDs)]{ssd}{SSD}{sequential social dilemma}

\begin{document}

\settopmatter{printacmref=false} 
\renewcommand\footnotetextcopyrightpermission[1]{} 
\pagestyle{plain} 


\pagestyle{fancy}
\fancyhead{}


\maketitle


\section{Introduction}

The standard approach to developing a \gls{rl} agent is to learn some fixed behaviour that will allow the agent to solve a sequential decision making problem. If, however, the developer wants the agent to behave differently, the agent normally has to be partially or completely retrained. To address this shortcoming, \citet{kallstrom2019tunable} introduced a framework to train agents whose behaviour can be tuned during run-time using \gls{morl} methods. In this framework, each set of objective preferences (scalarisation weights) corresponds to different combinations of desired agent behaviours, and the agent is trained with different weight vectors to learn different behaviours simultaneously. After the agent is trained, the weights can be adjusted on the fly to dynamically change the agent's behaviour, without the need for retraining. In this study we build on this framework, extending it to more complex environments with larger state-spaces and multiple learning agents. \par

In particular, we are interested in studying the suitability of the tunable agents framework to learn adaptive agent behaviours in \glspl{ssd} \cite{leibo2017multi}, settings where there is an inherent conflict between individual and collective welfare. In such settings, agents may choose to work together (cooperate) and share the resources available in the environment, or be selfish (compete) and attempt to maximise their own share of the resources, without considering the welfare of the other agents.

Agents that have learned a single fixed policy (as per standard RL methods) are at a significant disadvantage if the opponents in the environment change their strategies once training is finished. If an agent is trained to have a fixed cooperative policy, there is a chance that a competitive opponent may be able to exploit this, leaving the fixed cooperative agent with a poor payoff. Conversely, if an agent is trained to have a fixed competitive policy, it may miss out on greater rewards that might be possible by cooperating with a cooperative opponent. In either case, once training is complete, an agent trained using a standard single policy \gls{rl} approach is incapable of adapting its behaviour to changing environmental conditions without retraining, wasting valuable time and resources.

Our aim in this study is to establish how effective the tunable agents framework first introduced by \citet{kallstrom2019tunable} is for developing agents that are capable of adjusting their degree of cooperation in \glspl{ssd}. To this end, we conduct experiments in a modified version of the Wolfpack environment \cite{leibo2017multi}, an \gls{ssd} where multiple predator agents aim to capture a prey.
Our empirical results demonstrate that it is possible to train a single tunable agent that allows easy adaption between cooperative and competitive behaviours in sequential social dilemmas without the need for retraining, catering for a wide range of possible behaviours and opponent strategies. 


\section{Background}

\subsection{Multi-Agent Reinforcement Learning}

\gls{rl} is a machine learning paradigm that is concerned with enabling agents to learn how to solve sequential decision-making problems. The agent learns by receiving a reward after performing an action in an environment that represents how good or bad the action was \citep{kaelbling1996reinforcement}. Sequential decision-making problems are most commonly modelled as \glspl{mdp}. \glspl{mdp} consist of a set of environment states $S$, a set of actions that the agent can take $A$, a transition function $T$ and a reward function $R$. An \gls{mdp} is therefore defined as the tuple $\langle S,A,T,R \rangle$; if $s \in S$ is the current state of the agent, and it takes action $a \in A$, it will transition to state $s' \in S$ with probability $T(s,a,s') \in [0,1]$ and receive a real-valued reward $r=R(s,a,s')$ \citep{sutton2018reinforcement}.

An agent decides which action to take based on its policy $\pi$, which is effectively a mapping from environment states to agent actions. The goal of an agent in an \gls{mdp} is to find the policy that maximises its expected return at each time-step $t$; the return is defined as $g_t=\sum_{k=0}^{\infty} \gamma^{k} r_{t+k+1}$ where $\gamma \in [0,1]$ is the discount factor that controls how much the agent values future rewards. Value-based \gls{rl} algorithms are a common approach for solving \glspl{mdp}. These methods involve quantifying how good a particular state is using a value function. The value of a state when following policy $\pi$ is $V^\pi (s)=\mathbb{E}[g_t|\pi,s_t=s]$. Value-based methods therefore try to find the optimal policy $\pi^*=\argmax_{\pi}V^{\pi}(s)$ \citep{sutton2018reinforcement}.

A popular value-based \gls{rl} method is the Q-learning algorithm, which is a model-free, off-policy learning algorithm that estimates the action-value function $Q(s,a)$ of an \gls{mdp} by applying the following update rule at time-step $t$:

\begin{equation}
Q(s_t,a_t) \leftarrow Q(s_t,a_t) + \alpha \left( r_t + \gamma \max_{a} Q(s_{t+1},a) - Q(s_t,a_t) \right) 
\label{eqn:qlearning}
\end{equation}
\noindent where $\alpha$ is the learning rate and the value function is computed by $V(s)=\max_{a} Q(s,a)$ \citep{watkins1989learning}. Q-learning is a tabular method because the action-values for each state-action pair need to be stored in a table known as a Q-table. Tabular methods are normally only useful in very simple problems since the Q-table grows exponentially for every additional state dimension or action \citep{sutton2018reinforcement}. Therefore, Q-learning becomes infeasible for large state-spaces. To address this issue, methods that approximate the function $Q(s,a)$ are often used. One such method is the \gls{dqn} algorithm that involves training a deep neural network (referred to as a \gls{dqn}) to approximate the value function \citep{mnih2015human}.

Environments containing more than one agent are known as \glspl{mas}. \glspl{mas} are useful for solving problems that require more scalable and robust solutions as they are distributed by nature. Agents in an \gls{mas} may work cooperatively and/or competitively to solve a task \citep{wooldridge2009introduction}. \gls{marl} is an area of research in \gls{rl} that is concerned with solving problems using \gls{rl} techniques in multi-agent environments. \par

A \gls{sg} is a generalisation of the \gls{mdp} framework that enables the use of multiple agents. In an \gls{sg}, multiple agents perform actions at each time-step and the next state of the environment along with the reward for each agent are dependent on the joint actions of all of the agents \citep{busoniu2008comprehensive}. \par

An \gls{sg} between $n$ agents, consists of a set of states $S$, a set of actions for each agent $A_{1}, A_{2}, ..., A_{n}$, a transition function $T$ and a reward function for each agent $R_{1}, R_{2}, ..., R_{n}$ \citep{busoniu2008comprehensive}.  When agents have their own local state observation $o_i$, a degree of uncertainty over the environment state $s$ is introduced and the \gls{sg} is referred to as partially observable. An observation function $O$ is added to the definition of the \gls{sg}, which is used to compute the local state for each agent $i$ as $o_{i}=O(s,i)$ \citep{emery2004approximate}. \par

One approach to solving \glspl{sg} is to train each agent in the \gls{mas} using single-agent \gls{rl} techniques (such as the \gls{dqn} algorithm) and treat other agents as part of the environment \citep{mannion2017policy}.\par

\subsection{Multi-Objective Reinforcement Learning}
\label{sec:morl}

Standard \gls{rl} methods operate under the assumption that the problem can be solved by optimising a single objective; this is captured by the scalar reward received at each time-step. However, many real-world problems are multi-objective by nature and these objectives can be in conflict with each other \citep{vamplew2018human}. For example, if designing an agent to control a self-driving car, driving fast and keeping fuel consumption low would be conflicting objectives. \gls{morl} can be used to develop agents that can manage the trade-offs inherent in such problems. \par

For multi-objective sequential decision making problems, the \gls{mdp} framework needs to be extended to a \gls{momdp}, where at each time-step, the agent receives a vector of real-valued rewards $\textbf{r}_t$ (one reward for each objective). The vectorised value function is therefore defined as $\textbf{V}^\pi (s)=\mathbb{E}[\textbf{g}_t|\pi,s_t=s]$ where $\textbf{g}_t=\sum_{k=0}^{\infty} \gamma^{k} \textbf{r}_{t+k+1}$ \citep{roijers2013survey}. Note the bold font signifying vectors. For multi-objective multi-agent sequential decision making problems, the corresponding framework is a \gls{mosg}, where each agent has an individual reward function that returns reward vectors \cite{radulescu2020multi}.\par

Solution methods for \glspl{momdp} can be divided into two categories: single policy methods and multiple-policy methods \citep{vamplew2011empirical}. There is no single optimal policy to be found for \glspl{momdp}, in the same way that there is no single solution to a multi-objective optimisation problem. Instead, the solution is the set of points that form the Pareto front \citep{deb2014multi}. However, a scalarisation (or utility) function can be used to convert the output of the vectorised value function of the \gls{momdp} into a scalar, making it possible to find a single policy that maximises that utility function for a given set of scalarisation weights. The general form of a scalarisation function is $V_\textbf{w}^{\pi}(s)=f(\textbf{V}^{\pi}(s),\textbf{w})$ where $\textbf{w}$ is a vector of weights corresponding to the preferences between objectives. A common form of scalarisation applied in \gls{morl} is linear scalarisation: $V_\textbf{w}^{\pi}(s)=\textbf{w} \cdot \textbf{V}^{\pi}(s)$ \citep{roijers2013survey}. Although this method is widely used, it has the drawback that it cannot find any policy in a concave region of the Pareto front \citep{vamplew2008limitations}. If $\textbf{w}$ is unknown prior to the learning stage, a set of policies can be learned using a range of values for $\textbf{w}$ \citep{roijers2013survey}.


\section{Related Work}
Agents are most commonly designed to achieve some fixed goal behaviour. If a different behaviour is desired, the agent may need to be partially or completely redesigned or, in the case of an \gls{rl} agent, retrained. The ability to tune the behaviour of an agent in real time would be beneficial for many applications; for example, a stock trading agent could be tuned to take more risks depending on a customer's investment preferences, or video game playing agents could be tuned to be more aggressive or more defensive. \par

One approach to achieve this goal is to train an agent using different reward functions to yield several fixed policies with desired behaviours, and then switch between the policies during run-time to change the behaviour of the agent. Such an approach was adopted by \citet{klinkhammer2018learning} who studied how to select the best policy to follow at a given time based on the degree of alignment of sub-rewards with a global reward signal. \par

Another approach to designing agents capable of dynamic behaviours is to use concepts from \gls{morl}. A set of weights in the scalarisation function can be used to specify preferences for each type of desired behaviour, and after training an agent using the \gls{momdp} framework, the weights can be adjusted during run-time to switch between the behaviours. This idea was introduced by \citet{kallstrom2019tunable}. In this framework an agent receives a linearly scalarised reward based on objective preferences (weights) at each time-step.

\begin{figure}
  \centering
  \centerline{\includegraphics[width=0.8\linewidth]{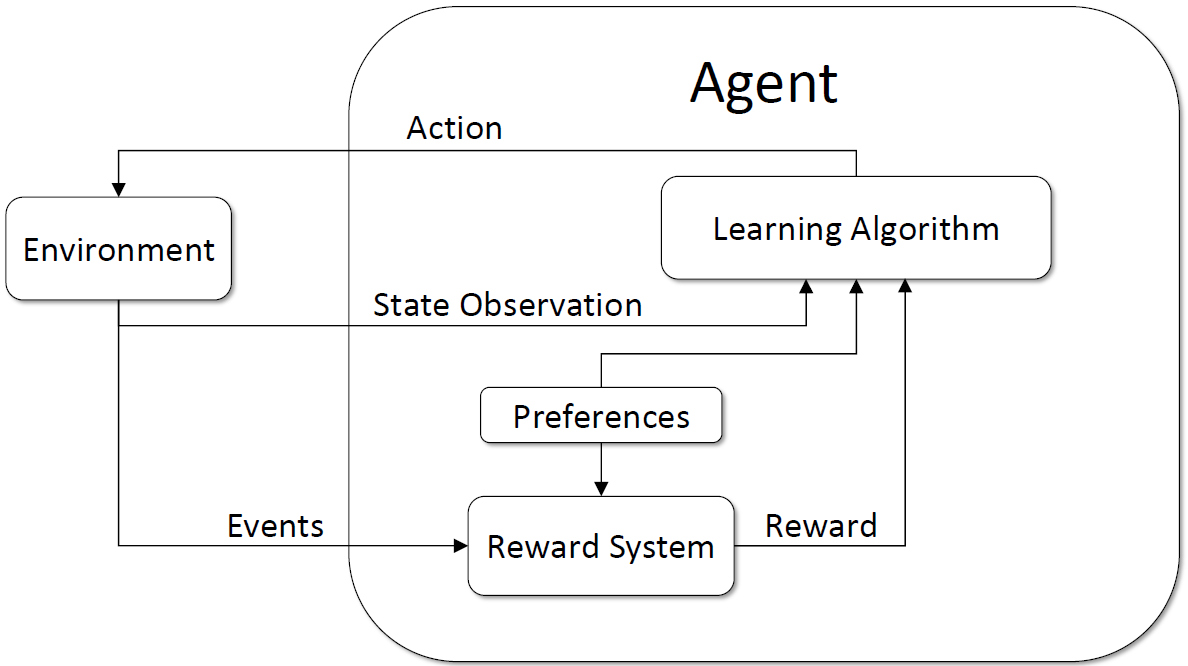}}
  \caption[Architecture for training agents with tunable behaviours]{Architecture for training agents with tunable behaviours. Image reproduced from \citet{kallstrom2019tunable}}
  \label{fig:kallstrom_framework}
\end{figure}

\citet{kallstrom2019tunable} used an adapted version of the \gls{dqn} algorithm to train the tunable agent; an important addition however is that the objective preferences (weights) are fed into the network along with the current state of the environment. This key modification means that the tunable \gls{dqn} agent can take account of its current preferences when predicting the value of taking an action, and that changing the behaviour of a tunable \gls{dqn} agent simply requires changing the current preference weights, avoiding the need for retraining. A block diagram of this framework is shown in Figure \ref{fig:kallstrom_framework}. The authors conducted two gridworld experiments to evaluate the framework and demonstrated empirically that the behaviours of agents could be tuned during run-time to behave similarly to agents trained using standard \gls{dqn} to reach fixed behaviours. The types of behaviours considered included cooperativeness, competitiveness and risk-aversion.

A further study was conducted using the tunable \gls{dqn} framework in a different environment by \citet{kallstrom2019multi}. One of the experiments in this second study focused on designing an agent for fighter-pilot training simulations that could be tuned to obtain a balance between behaviours that favour increasing safety or reducing time-to-destination. This work also demonstrated that tunable agents could reach similar behaviours when compared to agents trained using fixed preferences.


\section{Methodology}
\label{sec:methodology}
\subsection{Neural Network Architecture}
\label{sec:nn_arch}

In this work we use gridworld environments where the state is represented as a 3-channel RGB image showing the positions of items on the grid. Since the environments are multi-objective, the reward returned after each time-step is a 1-dimensional vector of a fixed length. An agent expresses its preferences over the elements in the reward vector through an objective preference weight vector (i.e., the scalarisation weights). Since the goal is to train an agent with tunable behaviours, the value predicted for each action needs to depend on both the current state of the environment and the objective preference weight vector. \par

The environment state observed by the agent at each time-step is comprised of the previous three frames of the environment stacked together. This in effect means that the \textit{current state} of the environment is a 9-channel image as opposed to a 3-channel one. Note that for the first two time-steps of each episode, this stack of frames includes duplicates to keep the size of the stack fixed. \par

\begin{figure}
  \centering
  \centerline{\includegraphics[width=\linewidth]{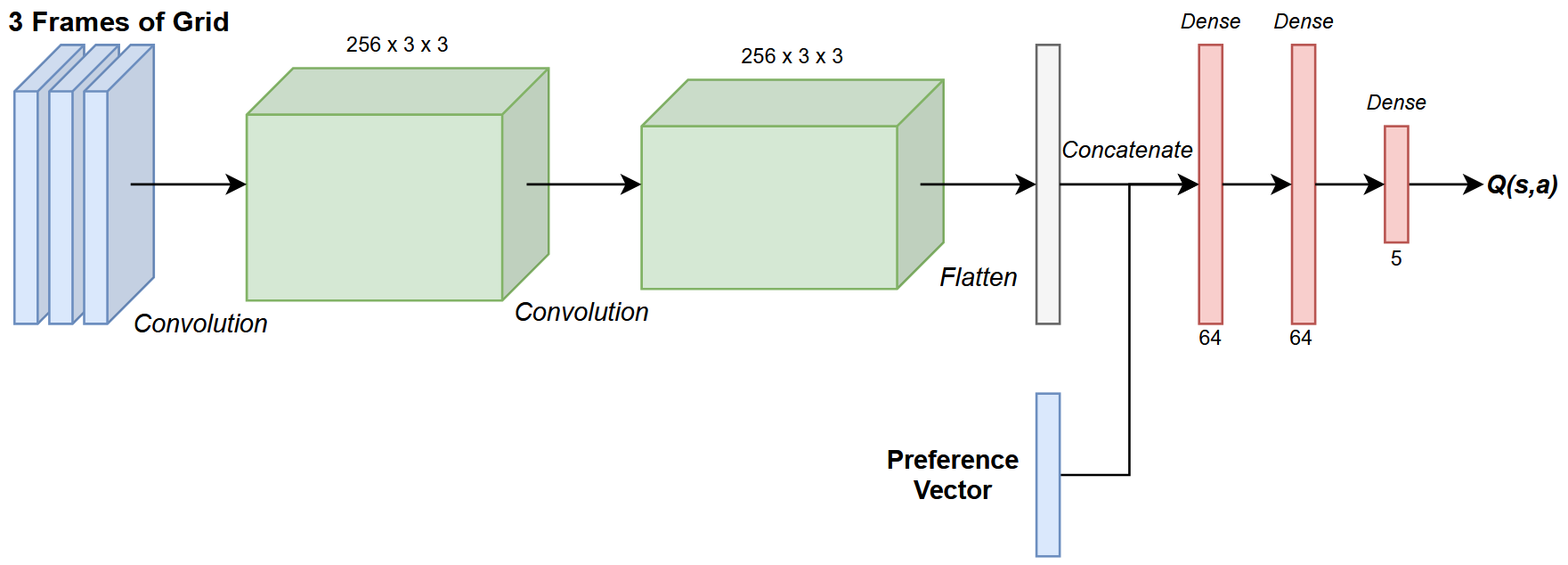}}
  \caption{Architecture of the deep neural network created for training tunable agents in our work.}
  \label{fig:tunable_dqn}
\end{figure}

The image pixel values are rescaled between 0 and 1 before being processed by two convolutional layers to extract features from the grid image. No form of pooling operation is used after each convolutional layer due to the image already being of relatively low dimensionality ($16 \times 16 \times 3$ per frame for Wolfpack). The output of the second convolutional layer is flattened to a 1-dimensional vector and concatenated with the preference weight vector, which is also rescaled between 0 and 1. The concatenated 1-dimensional vector is then passed through two fully-connected (i.e., dense) hidden layers and finally a dense output layer. \par

A block diagram of this architecture is shown in Figure \ref{fig:tunable_dqn}. Note that the dimensions of the convolutional layers shown refers to 256 filters of size $3 \times 3$. The \gls{relu} activation function is used after all convolutional layers and dense layers except for the output layer, which has a linear activation function since predicting Q-values is a regression problem. \par

Although the exact architecture used by \citet{kallstrom2019tunable} is not provided in their paper, the architecture described above was chosen to fit the description they provide as closely as possible. Certain specifications used in this work such as how to scale the data, the number of units in each layer and the filter sizes were found through experimentation as they were not reported by \citet{kallstrom2019tunable}. \par

\subsection{Tunable Agent Algorithm}

Algorithm \ref{alg:tunable_agents} outlines the method used for training the neural network described in Section \ref{sec:nn_arch}. The \gls{dqn} algorithm is a central component to this training scheme. \citet{kallstrom2019tunable} presented a high-level method for training tunable agents; Algorithm \ref{alg:tunable_agents} is a lower-level version of this method with steps specific to using the \gls{dqn} algorithm as the base \gls{rl} algorithm. \par 

At the beginning of each episode, the agent samples a set of objective preference weights from the preference sample space; this forms the objective preference weight vector $\textbf{w}$. The action at each time-step is chosen based on $\textbf{w}$ and the current state of the environment $s$ (the stack of the last three frames of the grid) using the network described in Section \ref{sec:nn_arch} with an $\epsilon$-greedy policy. After an action $a$ is executed, the next state of the environment $s'$ and reward vector $\textbf{r}$ is received from the environment. A scalar reward $r$ is then computed using linear scalarisation between $\textbf{r}$ and $\textbf{w}$. The agent then stores the experience tuple $(s, a, r, s', \textbf{w})$ in its replay memory. After each episode (excluding the first $T$ episodes to allow the replay memory to grow initially), the agent is trained on a minibatch from the replay memory. \par

A key difference to highlight between the single-objective \gls{dqn} algorithm and Algorithm \ref{alg:tunable_agents} is that the network is conditioned on the objective preference weight vector so \textbf{w} needs to also be stored in the experience replay memory. The training frequency is another difference to highlight. Here, the weights of the network are updated at the end of every episode, while in the \gls{dqn} algorithm, they are updated after every time-step. The training method described in \citet{kallstrom2019tunable} performs training steps less frequently (every $n$ episodes).

\begin{algorithm}
\SetAlgoLined

 Initialise $\epsilon$ to 1.0

 Set $T$ to the number of episodes to start training after 

\For{$episode\leftarrow 1$ \KwTo $M$}{
    \textit{Perform each command individually for agent $i=1,...,n$ where $n$ is the number of agents}
    
    Get the initial state $s_i$ from the environment

    Sample a set for preference weight vector ($\textbf{w}_i$) from the preference weight sample space

    \While{$s_i$ {\normalfont\textbf{is not}} terminal}{
        Choose action $a_i$ from state $s$ using $\epsilon$-greedy policy: $a_i=\texttt{agent[}i\texttt{].get\_action}(s_i,\textbf{w}_i,\epsilon)$

        Take action $a_i$ and observe reward vector $\textbf{r}_i$ and next state $s'_i$

        Store experience in replay memory: $\texttt{memory[}i\texttt{].append}(s_i,a_i,\textbf{r}_i,s'_i,\textbf{w}_i)$

        $s_i \leftarrow s'_i$
    }

    Decay $\epsilon$

    \If{$episode>T$}{
        $minibatch \leftarrow \texttt{memory[}i\texttt{].sample}()$

        $\texttt{agent[}i\texttt{].train}(minibatch)$
    }
}

 \caption{The tunable \gls{dqn} agent algorithm used in our work (adapted from \citet{kallstrom2019tunable})}
 \label{alg:tunable_agents}
\end{algorithm}

\section{Wolfpack Experiments}
\label{sec:wolfpack}

\subsection{Simulation Methods}
\label{sec:wolfpack_sim_methods}
\begin{figure}
  \centering
  \centerline{\includegraphics[width=0.5\linewidth]{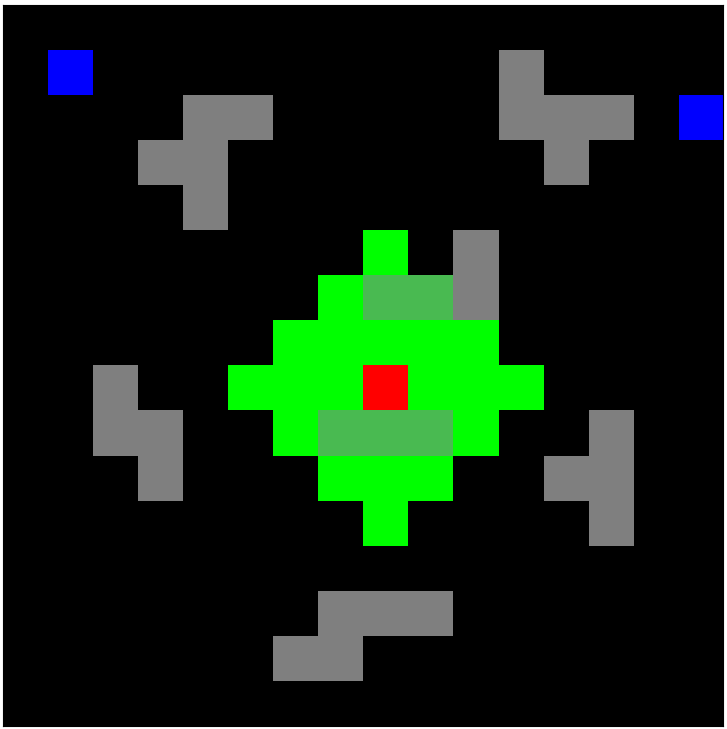}}
  \caption[Environment for the multi-objective Wolfpack experiment]{Environment for the Wolfpack experiment. Predators are represented by blue boxes, the prey is represented by a red box and the obstacles are the grey boxes. The capture-radius is also included and is shown in green.}
  \label{fig:wolfpack_grid}
\end{figure}

The Wolfpack environment involves three agents that can navigate around a gridworld-type space. There is one \textit{prey} agent and two \textit{predator}/\textit{wolf} agents. The predators must navigate around the grid to capture the prey, either as a pack (team) or alone. The grid that the agents can navigate is shown in Figure \ref{fig:wolfpack_grid}; it is of size $16 \times 16$ and contains several obstacles (shown in grey) that the agents must move around. The predators are shown in blue and the prey is shown in red. The prey is captured when one of the predators is at the same location as the prey. A team-capture occurs when the prey is captured by one predator while the other predator is within a certain radius, referred to as the capture-radius. The green area in Figure \ref{fig:wolfpack_grid} represents the capture-radius, which is only highlighted for the purpose of this figure and is not actually part of the grid image. The capture-radius was set to a Manhattan distance of 3 throughout this research. At each time-step, each agent can either remain in its current position or move up, down, left or right. \par

The Wolfpack environment used in the original study by \citet{leibo2017multi} was of a higher degree of complexity. The grid dimensions were $20 \times 20$, the agents had only partial observability of the grid that was dependent on their forward-facing direction and they could rotate as an action to change their direction of view. The environment was simplified in this study to reduce the training time for the agents. \par

The environment in the original study was also a single-objective stochastic game, with a different scalar reward received for team-captures and lone-captures. It was adapted to a multi-objective stochastic game for this research by giving four different reward signals at each time-step: \par

\begin{enumerate}
    \setlength\itemsep{0em}
    \item \textit{step}: A reward of $-1$ for each time-step to encourage the agents to complete an episode quickly.
    \item \textit{wall}: A reward of $-1$ any time the associated agent hits the grid boundary or an obstacle. 
    \item \textit{lone-capture}: A reward of $1$ if the associated agent captures the prey with no other predator inside the capture-radius.
    \item \textit{team-capture}: A reward of $1$ if the associated agent is inside the capture-radius with another predator when the prey is captured.
\end{enumerate}

At the beginning of each episode, the starting positions of the three agents are set randomly to empty locations in the grid. A reward vector is returned for each agent in the environment after each time-step and an episode ends if the prey gets captured or 150 time-steps have elapsed. A separate state for each agent is also returned after each time-step; in the image of the grid that each predator sees, they are represented as a blue box and the other predator is represented as a green box. This was done to allow the predator to learn to identify its own location in the grid and the other agents are just treated as part of the environment. A similar approach was taken by \citet{leibo2017multi}.

\begin{table}[H]
    \centering
    \caption{Hyperparameters for training the predator agents in the Wolfpack environment}
    \label{tab:wolfpack_hyperparams}
    \begin{tabular}{ll}
        \hline
        \textbf{Hyperparameter}                 & \textbf{Value} \\ \hline
        Loss Function                           & Huber          \\
        Optimizer                               & Adam           \\
        Learning Rate                           & 0.0001         \\
        Discount Factor                         & 0.99           \\
        Initial Epsilon                         & 1.0            \\
        Epsilon Decay                           & 1/21,250       \\
        Final Epsilon                           & 0.01           \\
        Replay Memory Size                      & 6,000          \\
        Minibatch Size                          & 64             \\
        Start training model after   (episodes) & 50             \\
        Copy to target every (steps)            & 1,000          \\
        Number of Training Episodes             & 80,000         \\ \hline
    \end{tabular}
\end{table}

For this study, the prey was not trained to evade the predators and it was just assigned a random-action policy. It is not clear what approach was taken by \citet{leibo2017multi} in this regard. The predators were trained as two separate tunable agents using Algorithm \ref{alg:tunable_agents}.  The agents were given a fixed preference of $0.005$ for \textit{step} and $0.025$ for \textit{wall} because there was no reason to have tunable preferences over these rewards. The \textit{lone-capture} preference was then sampled from 5 evenly spaced values between $0$ and $0.97$ and the \textit{team-capture} preference was chosen to make the full objective preference weight vector sum to 1. \par 

\begin{figure}
  \centering
  \centerline{\includegraphics[width=\linewidth]{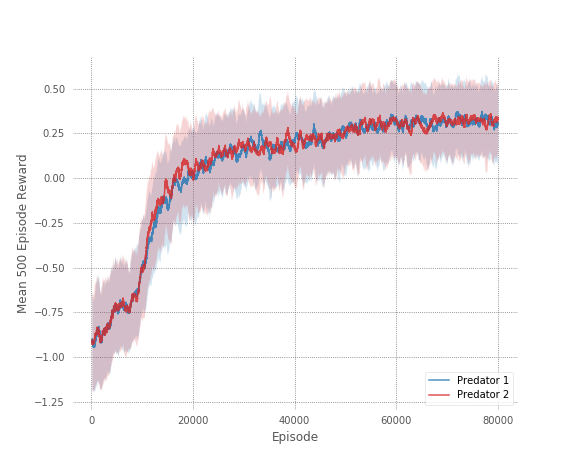}}
  \caption[Training progress for the two tunable predator agents in the Wolfpack environment]{Training progress for the two tunable predator agents in the Wolfpack environment. The shaded regions represent the error that is computed as the standard deviation.}
  \label{fig:wolfpack_training_progress}
\end{figure}

The hyperparameters used for training both tunable predators are shown in Table \ref{tab:wolfpack_hyperparams}. A 20\% dropout was also used in each convolutional layer to help prevent overfitting. The training progress plots for the two tunable predators are shown in Figure \ref{fig:wolfpack_training_progress}.

Agents with fixed preferences were also trained in the Wolfpack environment. To give the fixed agent experience of games with agents of different behaviours, a tunable agent was instantiated as the second predator during training time. This would allow a fair comparison between tunable and fixed agents. Two types of behaviours were considered for fixed agents:

\begin{enumerate}
    \setlength\itemsep{0em}
    \item \textit{Cooperative} : $\textbf{w}=\left[\begin{matrix} 0.005 & 0.025 & 0.0 &  0.97 \end{matrix}\right]$
    \item \textit{Competitive} / \textit{Defective} : $\textbf{w}=\left[\begin{matrix} 0.005 & 0.025 & 0.97 & 0.0 \end{matrix}\right]$
\end{enumerate}

\begin{figure}
  \centering
  \centerline{\includegraphics[width=\linewidth]{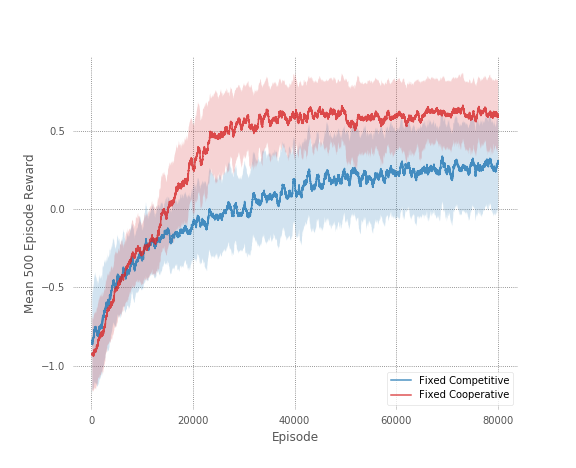}}
  \caption[Training progress for the two types of fixed behaviour predator agents in the Wolfpack environment]{Training progress for the two types of fixed behaviour predator agents in the Wolfpack environment. The shaded regions represent the error that is computed as the standard deviation.}
  \label{fig:wolfpack_training_progress_fixed}
\end{figure}

The training progress plots for the two types of fixed agents are shown in Figure \ref{fig:wolfpack_training_progress_fixed}.

The source code to reproduce our experiments may be downloaded from \url{https://github.com/docallaghan/tunable-agents}.

\subsection{Results}

\begin{figure}
  \centering
  \centerline{\includegraphics[width=\linewidth]{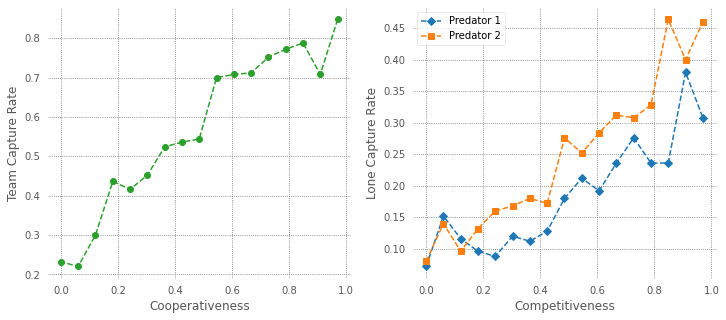}}
  \caption{Tuning performance for two predator agents with matched preferences}
  \label{fig:wolfpack_tuning_2pred}
\end{figure}

Results were collected to analyse the ability of the tunable agents to achieve competitive and cooperative behaviours of a varying degree. The first test was to instantiate two trained tunable predators in the Wolfpack environment and simulate 250 episodes for a range of different preference weights where the two predators had the same preference in any given episode. Figure \ref{fig:wolfpack_tuning_2pred} shows how the lone-capture rate and team-capture rates vary with the agents' degree of cooperativeness (team-capture preference) or competitiveness (lone-capture preference). \par

\begin{figure}
  \centering
  \centerline{\includegraphics[width=\linewidth]{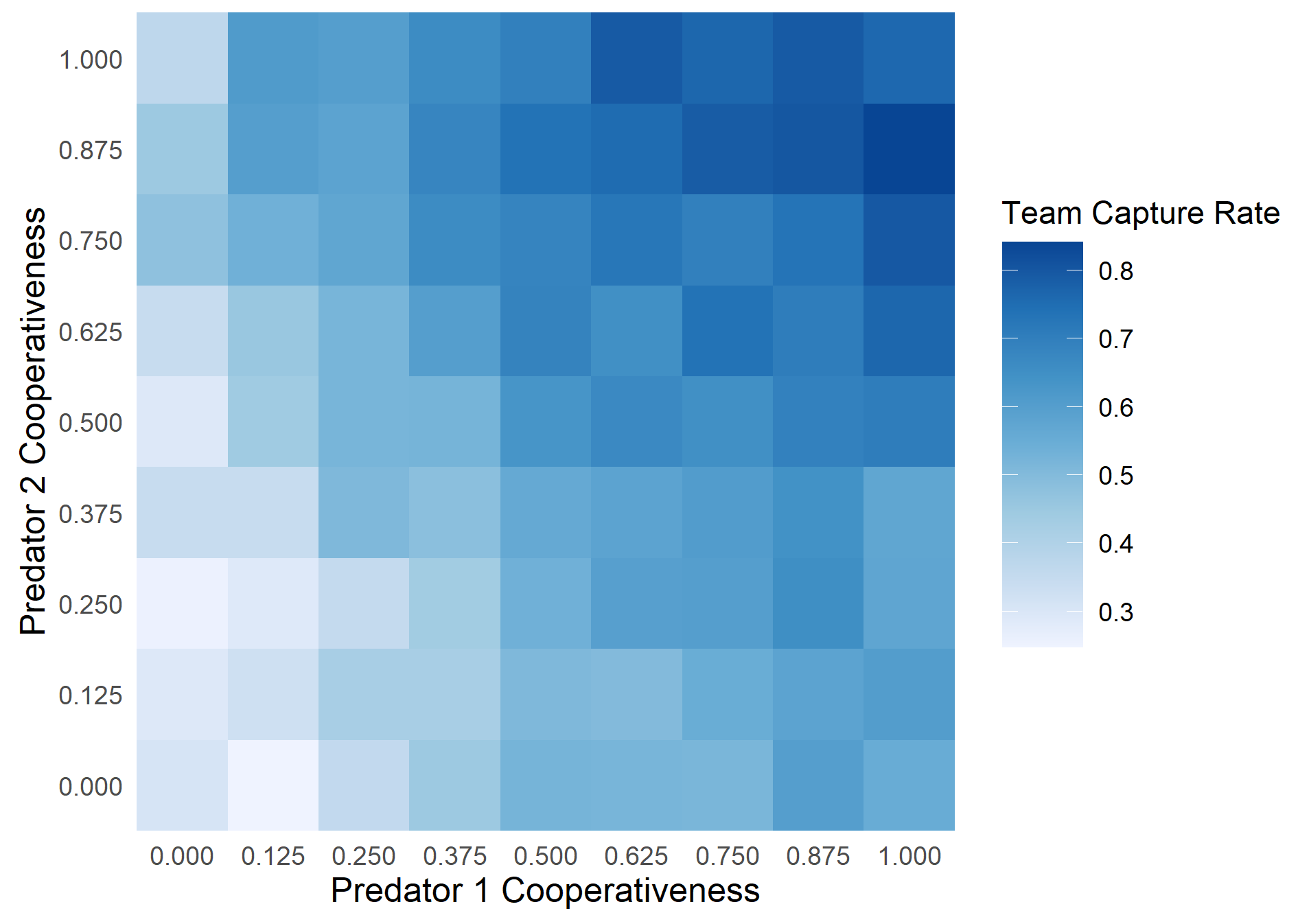}}
  \caption{Tuning performance for two predator agents with varied preferences}
  \label{fig:wolfpack_heatmap}
\end{figure}

Naturally, the next scenario to simulate was a series of encounters between tunable predators of varying preferences. Once again, 250 episodes were simulated of each game and the two different types of captures were tracked. The results are displayed as a heatmap in Figure \ref{fig:wolfpack_heatmap}, showing how the team-capture rate varies with the degree of cooperativeness of each predator. \par

\begin{figure}
  \centering
  \centerline{\includegraphics[width=\linewidth]{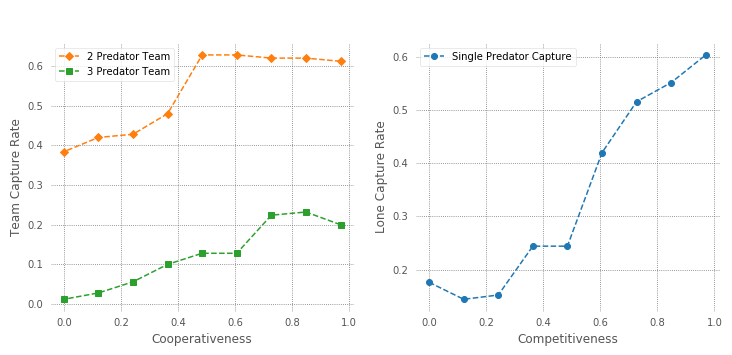}}
  \caption{Tuning performance for three predator agents with matched preferences}
  \label{fig:wolfpack_tuning_3pred}
\end{figure}

\begin{figure}
  \centering
  \centerline{\includegraphics[width=\linewidth]{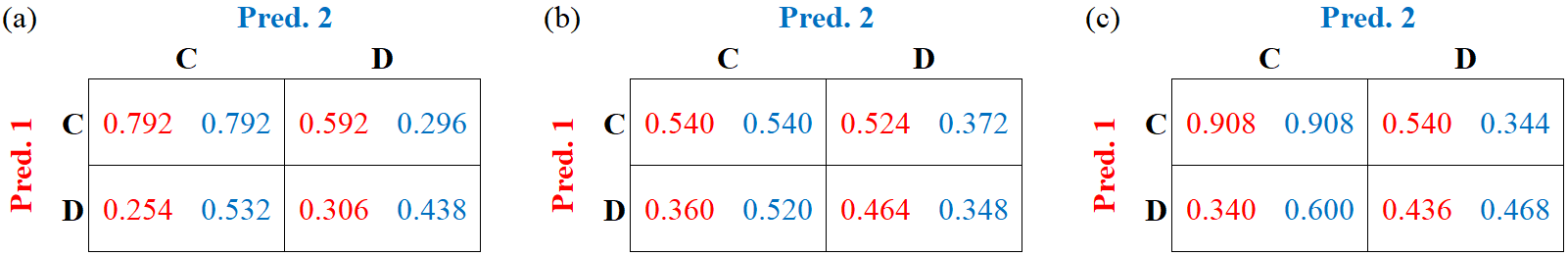}}
  \caption[Empirical payoff matrices for the Wolfpack experiment]{Empirical payoff matrices for the Wolfpack experiment. (a) Tunable agents (b) Fixed agents with matched models (c) Fixed agents with unique models.}
  \label{fig:wolfpack_payoff}
\end{figure}

Using a predator agent that was trained with a different random seed, a third predator was instantiated into the environment. The degree of cooperativeness and competitiveness was varied in simulations of 250 episodes for each setup (all three predators had matched preferences) and the team-capture and lone-capture rates were tracked. The results are shown in Figure \ref{fig:wolfpack_tuning_3pred}. \par

Empirical payoff matrices were generated by simulating 250 games (episodes) between fully cooperative agents and fully competitive (defective) agents (i.e., agents with the objective preferences weights used for training the fixed agents). The payoff for the encounter was computed as the capture rate (lone or team, depending on the type of agent) over the 250 episodes. Figure \ref{fig:wolfpack_payoff}a shows the empirical payoff matrix constructed for encounters between the two tunable predators whose training progress is shown in Figure \ref{fig:wolfpack_training_progress}. The equivalent payoff matrix for the two fixed agents is shown in Figure \ref{fig:wolfpack_payoff}b; this meant that, for the cooperative versus cooperative and the defective versus defective scenarios, the same model was instantiated twice. Figure \ref{fig:wolfpack_payoff}c shows the payoff matrix for fixed agents where two extra fixed agents were trained with different random seeds to avoid the scenario of two equivalent models playing against each other. The method for generating these empirical payoff matrices is based on the method used by \citet{leibo2017multi}. \par

\subsection{Discussion}
When viewing the behaviour of the tunable predators in simulation, the effects of varying the predators' objective preferences were clear. Setting a high preference for lone-captures made the predators more competitive and therefore less likely to participate in team captures. Once an episode began, both predators would move as quickly as possible to capture the prey on their own.

Conversely, setting a high preference for team-captures, made the predators more cooperative, greatly increasing the team capture rate as the cooperativeness preference was increased. At the beginning of each episode, both predators would typically move towards each other and approach the prey together, often taking different routes around obstacles to trap the prey \footnote{These two types of behaviours can be seen in the video recordings at \url{https://www.youtube.com/playlist?list=PLuIjfbXklqXbNY1tXl7gtEWj5J5pCH_Ub}}. \par

Referring to Figure \ref{fig:wolfpack_tuning_2pred}, a strong positive correlation between both tunable predators' level of cooperativeness and the resulting team-capture rate can be seen. The same can be said for both predators' level of competitiveness and their respective lone-capture rates. Note that during training, the tunable predators could only sample from 5 discrete preference weight vectors (as described in Section \ref{sec:wolfpack_sim_methods}); these plots therefore include data-points of simulations with several different preference weight vectors that are unseen to the predators and the models are still able to generalise to an appropriate level of cooperativeness or competitiveness. \par

From the simulations between tunable predators with independently varied objective preferences, it can be seen in Figure \ref{fig:wolfpack_heatmap} that the trend remains between individual predators' level of cooperativeness and the resulting team-capture rate. This is a very interesting result as it shows that the models don't learn any assumption of how the other predator behaves and that they generalise to encounters of any type of behaviour. This is a direct result of sampling the preference weight vector for each predator individually at the beginning of each episode during training as opposed to using the same preference for each one. Note that the plot also contains preference weights that were not used during training, once again highlighting the generalisability of the models. \par

As an additional experiment, a third predator was added into the Wolfpack environment was a test to see how well the models can generalise to unseen states. During training, the images of the grid that either of the networks see, contains one red box (for the prey), one blue box (for the predator that the network is being trained for) and only one green box (for the other predator). However, adding a third predator during simulation means that an image would contain two green boxes and that the state would not have been seen by the models during training. The plots in Figure \ref{fig:wolfpack_tuning_3pred}, however, show that the models can generalise to working in a 3-predator team and the level of cooperativeness can still be increased by tuning the objective preference weights. \par

Referring to the payoff matrix in Figure \ref{fig:wolfpack_payoff}a, it can be seen that the rational behaviour is to cooperate in any encounter since the payoff for cooperating is highest no matter what type of behaviour the opposing predator possesses. Note that there is no element of greed (when the temptation payoff is greater than the reward payoff) or fear (when the punishment payoff is greater than the sucker payoff) in this matrix game. Therefore, this game does not meet the conditions for a social dilemma that are outlined in \citet{macy2002learning}. In the version of the Wolfpack experiment by \citet{leibo2017multi}, empirical payoff matrices that were not social dilemmas were also seen for certain reward structures. Payoff matrices could have been generated for agents of different levels of cooperativeness and perhaps different social dynamics would have been seen since this would be analogous to changing the reward structure. However, this was not carried out since the main reason payoff matrices were generated in this study was to compare the performance of tunable and fixed preference agents as opposed to analysing the social dynamics. \par

The payoff matrix generated for the two fixed agents (Figure \ref{fig:wolfpack_payoff}b) yielded some unexpected results; the payoff for mutual cooperation is significantly lower than it was for the tunable agents. This suggests that the tunable agents reached higher levels of performance over fixed agents, going against what was seen in \citet{kallstrom2019tunable}. However, from viewing simulations of episodes of mutual cooperation, it was discovered that the reason for the low payoff was that both predators frequently met at the same grid location (making only one predator visible) and would then take the same actions until the end of the episode as both used the same \gls{dnn} model. Since only one predator was visible to each predator, it appeared as though there was no other predator present to cooperate with and therefore no incentive to approach the prey. The fixed competitive agents, however, didn't tend to approach each other so this was not an issue for that type of encounter. This was the reason for training two extra fixed agents using different random seeds. The resulting payoff matrix in Figure \ref{fig:wolfpack_payoff}c shows this behaviour disappear and dynamics closer to that of Figure \ref{fig:wolfpack_payoff}a are seen. \par

The values in the payoff matrices in Figures \ref{fig:wolfpack_payoff}a and \ref{fig:wolfpack_payoff}c are indeed quite similar, showing that tunable agents can reach similar behaviours to fixed agents. The fixed agents do however have slightly higher capture rates in all scenarios compared to the tunable agents; this observation is in alignment with the findings by \citet{kallstrom2019tunable}, who reported slightly reduced performance for tunable agents compared to fixed agents when the same preferences are used. This tradeoff is to be expected, given that the tunable agent architecture can exhibit a much wider range of behaviours when compared to fixed preference agents. 


\section{Conclusion and Future Work}
\label{sec:conclusions}

Our aim in this study was to establish how effective the tunable agents framework first introduced by \citet{kallstrom2019tunable} is for developing agents that are capable of adjusting their degree of cooperation in \glspl{ssd}. To this end, we conducted experiments in a modified version of the Wolfpack environment \cite{leibo2017multi}, a \gls{ssd} where multiple predator agents aim to capture a prey.

This study has shown that the tunable agents framework first introduced by \citet{kallstrom2019tunable} can be used to train multiple agents that are capable of tunable levels of cooperation in \glspl{ssd}. 
The results in the Wolfpack domain demonstrated that objective preferences for either trained predator agent could be tuned independently to achieve varying degrees of cooperative behaviour, and that there is a strong correlation between agents' preferences for cooperative behaviour and their tendencies to participate in cooperative team captures during simulations. \par

A further contribution made by this work is that it was empirically shown that the trained tunable predator agents could generalise well to unseen objective preference weightings, as well as unseen environment states. The former was shown by collecting results for objective preference weightings that were not used during the training cycle for the tunable agents (see Figures \ref{fig:wolfpack_tuning_2pred} and \ref{fig:wolfpack_heatmap}). The latter was shown by instantiating a third pre-trained predator agent into the Wolfpack environment during simulation and seeing that tunable cooperativeness for all three predators could still be achieved (see Figure \ref{fig:wolfpack_tuning_3pred}). \par

The contributions of this work open the door for this method of training agents with tunable behaviours to be applied to a huge array of different problems. This framework would be beneficial to any \gls{rl} problem where there is some degree of uncertainty over the desired type of agent behaviour. If an agent with fixed objective preferences was trained and it was then seen that the behaviour needed to be changed slightly, the agent would need to be retrained with new objective preferences. However, using the method for training tunable agents that was the focus of this study, the objective preferences could simply be fine-tuned after training. \par

There are many possible directions that could be taken to extend this research in the future.
One potential avenue for future work is to research the effect of different exploration strategies on the tunable \gls{dqn} model. In this study and in the original work by \citet{kallstrom2019tunable} $\epsilon$-greedy exploration was used, and no other strategy was tested. It would be valuable to assess the impact of more sophisticated exploration strategies such as softmax exploration on the learning dynamics of the tunable \gls{dqn} model. \par

The experimentation in the Wolfpack environment could be expanded on significantly in the future. The size of the grid could be increased from $16 \times 16$ and the agents could be given partial observability of the state-space, as was done in the original study by \citet{leibo2017multi}. Furthermore, the prey could also be trained to evade the predators; this could possibly introduce different social dynamics between the predators as it would be very difficult to capture the prey alone. \par

Another suggestion for future work is to apply this tunable agents framework to other more complex multi-agent environments, such as the \textit{Half Field Offence} 2D RoboCup Soccer simulation environment \citep{hausknecht2016half}. In this problem, agents could be trained to have tunable levels of attacking or defensive behaviours. Video games such as first person shooters, role-playing games or real-time strategy games are other exciting potential application areas, where non-player characters could be dynamically tuned to be more aggressive or defensive, or more or less economical with scarce resources such as food, ammunition or metals. Deep RL for autonomous driving \cite{kiran2020deep} is another potential application area, where users' preferences for factors such as fuel economy, comfort and route choices could be adjusted in real time using this framework.\par

The base \gls{rl} algorithm used for training tunable agents is another aspect of this research that could be investigated further. The training scheme used in this work (see Section \ref{sec:methodology}) could easily be adapted to use a base \gls{rl} algorithm other than the \gls{dqn} algorithm. One suggestion is to adapt the training scheme to work with actor-critic methods as they have been very successful in deep \gls{rl} applications in recent years. \par

In this work, linear scalarisation was used to compute the scalarised rewards when training the tunable agents. This same approach was taken by \citet{kallstrom2019tunable}. As mentioned in Section \ref{sec:morl}, linear scalarisation in \gls{morl} has its limitations as it can not find policies in concave regions of the Pareto front. It would therefore be beneficial to investigate the possibility of adapting the training methods used in this study to work with some form of non-linear scalarisation, such as non-linear monotonically increasing scalarisation functions as discussed by \citet{roijers2013survey}. More complex non-linear scalarisation functions could potentially allow tunable agents' preferences over behaviours to be represented in a more nuanced manner, and would also potentially fit better with how utility is derived in real-world multi-objective decision making problems (e.g. utility functions are non-linear in situations where a minimum value must achieved on each objective).\par

In our experiments, agent preferences were set manually ahead of each episode. Extending the tunable agents framework to allow agents to automatically classify opponents and modify their preferences accordingly during run-time is an interesting future research direction; to successfully achieve this it is likely that some form of opponent modelling for multi-objective settings (e.g. \cite{Zhang2020OpponentAAMAS,radulescu2020opponent}) would be necessary.

Finally, the impact of dynamically altering the preferences encoded in agents' utility functions should be analysed from a game theoretic perspective. Recent work analysed the impact of non-linear utility functions in multi-objective multi-agent settings \cite{radulescu2020utility}; however, this analysis was limited to fixed utility functions only, and should therefore be extended to take account of tunable agent preferences.

\bibliographystyle{ACM-Reference-Format} 
\bibliography{sample}


\begin{thebibliography}{25}


\ifx \showCODEN    \undefined \def \showCODEN     #1{\unskip}     \fi
\ifx \showDOI      \undefined \def \showDOI       #1{#1}\fi
\ifx \showISBNx    \undefined \def \showISBNx     #1{\unskip}     \fi
\ifx \showISBNxiii \undefined \def \showISBNxiii  #1{\unskip}     \fi
\ifx \showISSN     \undefined \def \showISSN      #1{\unskip}     \fi
\ifx \showLCCN     \undefined \def \showLCCN      #1{\unskip}     \fi
\ifx \shownote     \undefined \def \shownote      #1{#1}          \fi
\ifx \showarticletitle \undefined \def \showarticletitle #1{#1}   \fi
\ifx \showURL      \undefined \def \showURL       {\relax}        \fi
\providecommand\bibfield[2]{#2}
\providecommand\bibinfo[2]{#2}
\providecommand\natexlab[1]{#1}
\providecommand\showeprint[2][]{arXiv:#2}

\bibitem[\protect\citeauthoryear{Busoniu, Babuska, and De~Schutter}{Busoniu
  et~al\mbox{.}}{2008}]%
        {busoniu2008comprehensive}
\bibfield{author}{\bibinfo{person}{Lucian Busoniu}, \bibinfo{person}{Robert
  Babuska}, {and} \bibinfo{person}{Bart De~Schutter}.}
  \bibinfo{year}{2008}\natexlab{}.
\newblock \showarticletitle{A comprehensive survey of multiagent reinforcement
  learning}.
\newblock \bibinfo{journal}{\emph{IEEE Transactions on Systems, Man, and
  Cybernetics, Part C (Applications and Reviews)}} \bibinfo{volume}{38},
  \bibinfo{number}{2} (\bibinfo{year}{2008}), \bibinfo{pages}{156--172}.
\newblock


\bibitem[\protect\citeauthoryear{Deb}{Deb}{2014}]%
        {deb2014multi}
\bibfield{author}{\bibinfo{person}{Kalyanmoy Deb}.}
  \bibinfo{year}{2014}\natexlab{}.
\newblock \showarticletitle{Multi-objective optimization}.
\newblock In \bibinfo{booktitle}{\emph{Search methodologies}}.
  \bibinfo{publisher}{Springer}, \bibinfo{pages}{403--449}.
\newblock


\bibitem[\protect\citeauthoryear{Emery-Montemerlo, Gordon, Schneider, and
  Thrun}{Emery-Montemerlo et~al\mbox{.}}{2004}]%
        {emery2004approximate}
\bibfield{author}{\bibinfo{person}{Rosemary Emery-Montemerlo},
  \bibinfo{person}{Geoff Gordon}, \bibinfo{person}{Jeff Schneider}, {and}
  \bibinfo{person}{Sebastian Thrun}.} \bibinfo{year}{2004}\natexlab{}.
\newblock \showarticletitle{Approximate solutions for partially observable
  stochastic games with common payoffs}. In
  \bibinfo{booktitle}{\emph{Proceedings of the Third International Joint
  Conference on Autonomous Agents and Multiagent Systems, 2004. AAMAS 2004.}}
  IEEE, \bibinfo{pages}{136--143}.
\newblock


\bibitem[\protect\citeauthoryear{Hausknecht, Mupparaju, Subramanian,
  Kalyanakrishnan, and Stone}{Hausknecht et~al\mbox{.}}{2016}]%
        {hausknecht2016half}
\bibfield{author}{\bibinfo{person}{Matthew Hausknecht},
  \bibinfo{person}{Prannoy Mupparaju}, \bibinfo{person}{Sandeep Subramanian},
  \bibinfo{person}{Shivaram Kalyanakrishnan}, {and} \bibinfo{person}{Peter
  Stone}.} \bibinfo{year}{2016}\natexlab{}.
\newblock \showarticletitle{Half field offense: An environment for multiagent
  learning and ad hoc teamwork}. In \bibinfo{booktitle}{\emph{Adaptive and
  Learning Agents Workshop (ALA-19) at AAMAS, Montreal, Canada, May 13-14,
  2019}}.
\newblock


\bibitem[\protect\citeauthoryear{Kaelbling, Littman, and Moore}{Kaelbling
  et~al\mbox{.}}{1996}]%
        {kaelbling1996reinforcement}
\bibfield{author}{\bibinfo{person}{Leslie~Pack Kaelbling},
  \bibinfo{person}{Michael~L Littman}, {and} \bibinfo{person}{Andrew~W Moore}.}
  \bibinfo{year}{1996}\natexlab{}.
\newblock \showarticletitle{Reinforcement learning: A survey}.
\newblock \bibinfo{journal}{\emph{Journal of artificial intelligence research}}
   \bibinfo{volume}{4} (\bibinfo{year}{1996}), \bibinfo{pages}{237--285}.
\newblock


\bibitem[\protect\citeauthoryear{K{\"a}llstr{\"o}m and
  Heintz}{K{\"a}llstr{\"o}m and Heintz}{2019a}]%
        {kallstrom2019multi}
\bibfield{author}{\bibinfo{person}{Johan K{\"a}llstr{\"o}m} {and}
  \bibinfo{person}{Fredrik Heintz}.} \bibinfo{year}{2019}\natexlab{a}.
\newblock \showarticletitle{Multi-Agent Multi-Objective Deep Reinforcement
  Learning for Efficient and Effective Pilot Training}. In
  \bibinfo{booktitle}{\emph{FT2019. Proceedings of the 10th Aerospace
  Technology Congress, October 8-9, 2019, Stockholm, Sweden}}.
  \bibinfo{pages}{101--111}.
\newblock


\bibitem[\protect\citeauthoryear{K{\"a}llstr{\"o}m and
  Heintz}{K{\"a}llstr{\"o}m and Heintz}{2019b}]%
        {kallstrom2019tunable}
\bibfield{author}{\bibinfo{person}{Johan K{\"a}llstr{\"o}m} {and}
  \bibinfo{person}{Fredrik Heintz}.} \bibinfo{year}{2019}\natexlab{b}.
\newblock \showarticletitle{Tunable dynamics in agent-based simulation using
  multi-objective reinforcement learning}. In
  \bibinfo{booktitle}{\emph{Adaptive and Learning Agents Workshop (ALA-19) at
  AAMAS, Montreal, Canada, May 13-14, 2019}}. \bibinfo{pages}{1--7}.
\newblock


\bibitem[\protect\citeauthoryear{Kiran, Sobh, Talpaert, Mannion, Sallab,
  Yogamani, and P{\'e}rez}{Kiran et~al\mbox{.}}{2020}]%
        {kiran2020deep}
\bibfield{author}{\bibinfo{person}{B~Ravi Kiran}, \bibinfo{person}{Ibrahim
  Sobh}, \bibinfo{person}{Victor Talpaert}, \bibinfo{person}{Patrick Mannion},
  \bibinfo{person}{Ahmad A~Al Sallab}, \bibinfo{person}{Senthil Yogamani},
  {and} \bibinfo{person}{Patrick P{\'e}rez}.} \bibinfo{year}{2020}\natexlab{}.
\newblock \showarticletitle{Deep reinforcement learning for autonomous driving:
  A survey}.
\newblock \bibinfo{journal}{\emph{arXiv preprint arXiv:2002.00444}}
  (\bibinfo{year}{2020}).
\newblock


\bibitem[\protect\citeauthoryear{Klinkhammer, Yates, Tuladhar, and
  Tumer}{Klinkhammer et~al\mbox{.}}{2018}]%
        {klinkhammer2018learning}
\bibfield{author}{\bibinfo{person}{Eric Klinkhammer}, \bibinfo{person}{Connor
  Yates}, \bibinfo{person}{Yathartha Tuladhar}, {and} \bibinfo{person}{Kagan
  Tumer}.} \bibinfo{year}{2018}\natexlab{}.
\newblock \showarticletitle{Learning in Complex Domains: Leveraging Multiple
  Rewards through Alignment}. In \bibinfo{booktitle}{\emph{Adaptive and
  Learning Agents Workshop (ALA-18) at AAMAS, Stockholm, Sweden, July 14-15,
  2018}}. \bibinfo{pages}{1--9}.
\newblock


\bibitem[\protect\citeauthoryear{Leibo, Zambaldi, Lanctot, Marecki, and
  Graepel}{Leibo et~al\mbox{.}}{2017}]%
        {leibo2017multi}
\bibfield{author}{\bibinfo{person}{Joel~Z. Leibo}, \bibinfo{person}{Vinicius
  Zambaldi}, \bibinfo{person}{Marc Lanctot}, \bibinfo{person}{Janusz Marecki},
  {and} \bibinfo{person}{Thore Graepel}.} \bibinfo{year}{2017}\natexlab{}.
\newblock \showarticletitle{Multi-Agent Reinforcement Learning in Sequential
  Social Dilemmas}. In \bibinfo{booktitle}{\emph{Proceedings of the 16th
  Conference on Autonomous Agents and MultiAgent Systems}} (S\~{a}o Paulo,
  Brazil) \emph{(\bibinfo{series}{AAMAS '17})}.
  \bibinfo{publisher}{International Foundation for Autonomous Agents and
  Multiagent Systems}, \bibinfo{address}{Richland, SC},
  \bibinfo{pages}{464–473}.
\newblock


\bibitem[\protect\citeauthoryear{Macy and Flache}{Macy and Flache}{2002}]%
        {macy2002learning}
\bibfield{author}{\bibinfo{person}{Michael~W Macy} {and}
  \bibinfo{person}{Andreas Flache}.} \bibinfo{year}{2002}\natexlab{}.
\newblock \showarticletitle{Learning dynamics in social dilemmas}.
\newblock \bibinfo{journal}{\emph{Proceedings of the National Academy of
  Sciences}} \bibinfo{volume}{99}, \bibinfo{number}{suppl 3}
  (\bibinfo{year}{2002}), \bibinfo{pages}{7229--7236}.
\newblock


\bibitem[\protect\citeauthoryear{Mannion, Devlin, Mason, Duggan, and
  Howley}{Mannion et~al\mbox{.}}{2017}]%
        {mannion2017policy}
\bibfield{author}{\bibinfo{person}{Patrick Mannion}, \bibinfo{person}{Sam
  Devlin}, \bibinfo{person}{Karl Mason}, \bibinfo{person}{Jim Duggan}, {and}
  \bibinfo{person}{Enda Howley}.} \bibinfo{year}{2017}\natexlab{}.
\newblock \showarticletitle{Policy invariance under reward transformations for
  multi-objective reinforcement learning}.
\newblock \bibinfo{journal}{\emph{Neurocomputing}}  \bibinfo{volume}{263}
  (\bibinfo{year}{2017}), \bibinfo{pages}{60--73}.
\newblock


\bibitem[\protect\citeauthoryear{Mnih, Kavukcuoglu, Silver, Rusu, Veness,
  Bellemare, Graves, Riedmiller, Fidjeland, Ostrovski, et~al\mbox{.}}{Mnih
  et~al\mbox{.}}{2015}]%
        {mnih2015human}
\bibfield{author}{\bibinfo{person}{Volodymyr Mnih}, \bibinfo{person}{Koray
  Kavukcuoglu}, \bibinfo{person}{David Silver}, \bibinfo{person}{Andrei~A
  Rusu}, \bibinfo{person}{Joel Veness}, \bibinfo{person}{Marc~G Bellemare},
  \bibinfo{person}{Alex Graves}, \bibinfo{person}{Martin Riedmiller},
  \bibinfo{person}{Andreas~K Fidjeland}, \bibinfo{person}{Georg Ostrovski},
  {et~al\mbox{.}}} \bibinfo{year}{2015}\natexlab{}.
\newblock \showarticletitle{Human-level control through deep reinforcement
  learning}.
\newblock \bibinfo{journal}{\emph{Nature}} \bibinfo{volume}{518},
  \bibinfo{number}{7540} (\bibinfo{year}{2015}), \bibinfo{pages}{529--533}.
\newblock


\bibitem[\protect\citeauthoryear{O'Callaghan and Mannion}{O'Callaghan and
  Mannion}{2021}]%
        {ocallaghan2021tunable}
\bibfield{author}{\bibinfo{person}{David O'Callaghan} {and}
  \bibinfo{person}{Patrick Mannion}.} \bibinfo{year}{2021}\natexlab{}.
\newblock \showarticletitle{Tunable Behaviours in Sequential Social Dilemmas
  using Multi-Objective Reinforcement Learning}. In
  \bibinfo{booktitle}{\emph{Proceedings of the 20th International Conference on
  Autonomous Agents and Multiagent Systems (AAMAS 2021)}}.
\newblock


\bibitem[\protect\citeauthoryear{R{\u{a}}dulescu, Mannion, Roijers, and
  Now{\'e}}{R{\u{a}}dulescu et~al\mbox{.}}{2020a}]%
        {radulescu2020multi}
\bibfield{author}{\bibinfo{person}{Roxana R{\u{a}}dulescu},
  \bibinfo{person}{Patrick Mannion}, \bibinfo{person}{Diederik~M Roijers},
  {and} \bibinfo{person}{Ann Now{\'e}}.} \bibinfo{year}{2020}\natexlab{a}.
\newblock \showarticletitle{Multi-objective multi-agent decision making: a
  utility-based analysis and survey}.
\newblock \bibinfo{journal}{\emph{Autonomous Agents and Multi-Agent Systems}}
  \bibinfo{volume}{34}, \bibinfo{number}{1} (\bibinfo{year}{2020}),
  \bibinfo{pages}{10}.
\newblock


\bibitem[\protect\citeauthoryear{R{\u{a}}dulescu, Verstraeten, Zhang, Mannion,
  Roijers, and Now{\'e}}{R{\u{a}}dulescu et~al\mbox{.}}{2020b}]%
        {radulescu2020opponent}
\bibfield{author}{\bibinfo{person}{Roxana R{\u{a}}dulescu},
  \bibinfo{person}{Timothy Verstraeten}, \bibinfo{person}{Yijie Zhang},
  \bibinfo{person}{Patrick Mannion}, \bibinfo{person}{Diederik~M Roijers},
  {and} \bibinfo{person}{Ann Now{\'e}}.} \bibinfo{year}{2020}\natexlab{b}.
\newblock \showarticletitle{Opponent Learning Awareness and Modelling in
  Multi-Objective Normal Form Games}.
\newblock \bibinfo{journal}{\emph{arXiv preprint arXiv:2011.07290}}
  (\bibinfo{year}{2020}).
\newblock


\bibitem[\protect\citeauthoryear{Roijers, Vamplew, Whiteson, and
  Dazeley}{Roijers et~al\mbox{.}}{2013}]%
        {roijers2013survey}
\bibfield{author}{\bibinfo{person}{Diederik~M Roijers}, \bibinfo{person}{Peter
  Vamplew}, \bibinfo{person}{Shimon Whiteson}, {and} \bibinfo{person}{Richard
  Dazeley}.} \bibinfo{year}{2013}\natexlab{}.
\newblock \showarticletitle{A survey of multi-objective sequential
  decision-making}.
\newblock \bibinfo{journal}{\emph{Journal of Artificial Intelligence Research}}
   \bibinfo{volume}{48} (\bibinfo{year}{2013}), \bibinfo{pages}{67--113}.
\newblock


\bibitem[\protect\citeauthoryear{R\u{a}dulescu, Mannion, Zhang, Roijers, and
  Now\'{e}}{R\u{a}dulescu et~al\mbox{.}}{2020}]%
        {radulescu2020utility}
\bibfield{author}{\bibinfo{person}{Roxana R\u{a}dulescu},
  \bibinfo{person}{Patrick Mannion}, \bibinfo{person}{Yijie Zhang},
  \bibinfo{person}{Diederik~Marijn Roijers}, {and} \bibinfo{person}{Ann
  Now\'{e}}.} \bibinfo{year}{2020}\natexlab{}.
\newblock \showarticletitle{A utility-based analysis of equilibria in
  multi-objective normal form games}.
\newblock \bibinfo{journal}{\emph{The Knowledge Engineering Review}}
  \bibinfo{volume}{35}, \bibinfo{number}{e32} (\bibinfo{year}{2020}).
\newblock


\bibitem[\protect\citeauthoryear{Sutton and Barto}{Sutton and Barto}{2018}]%
        {sutton2018reinforcement}
\bibfield{author}{\bibinfo{person}{Richard~S Sutton} {and}
  \bibinfo{person}{Andrew~G Barto}.} \bibinfo{year}{2018}\natexlab{}.
\newblock \bibinfo{booktitle}{\emph{Reinforcement learning: An introduction}}.
\newblock \bibinfo{publisher}{MIT press}.
\newblock


\bibitem[\protect\citeauthoryear{Vamplew, Dazeley, Berry, Issabekov, and
  Dekker}{Vamplew et~al\mbox{.}}{2011}]%
        {vamplew2011empirical}
\bibfield{author}{\bibinfo{person}{Peter Vamplew}, \bibinfo{person}{Richard
  Dazeley}, \bibinfo{person}{Adam Berry}, \bibinfo{person}{Rustam Issabekov},
  {and} \bibinfo{person}{Evan Dekker}.} \bibinfo{year}{2011}\natexlab{}.
\newblock \showarticletitle{Empirical evaluation methods for multiobjective
  reinforcement learning algorithms}.
\newblock \bibinfo{journal}{\emph{Machine learning}} \bibinfo{volume}{84},
  \bibinfo{number}{1-2} (\bibinfo{year}{2011}), \bibinfo{pages}{51--80}.
\newblock


\bibitem[\protect\citeauthoryear{Vamplew, Dazeley, Foale, Firmin, and
  Mummery}{Vamplew et~al\mbox{.}}{2018}]%
        {vamplew2018human}
\bibfield{author}{\bibinfo{person}{Peter Vamplew}, \bibinfo{person}{Richard
  Dazeley}, \bibinfo{person}{Cameron Foale}, \bibinfo{person}{Sally Firmin},
  {and} \bibinfo{person}{Jane Mummery}.} \bibinfo{year}{2018}\natexlab{}.
\newblock \showarticletitle{Human-aligned artificial intelligence is a
  multiobjective problem}.
\newblock \bibinfo{journal}{\emph{Ethics and Information Technology}}
  \bibinfo{volume}{20}, \bibinfo{number}{1} (\bibinfo{year}{2018}),
  \bibinfo{pages}{27--40}.
\newblock


\bibitem[\protect\citeauthoryear{Vamplew, Yearwood, Dazeley, and Berry}{Vamplew
  et~al\mbox{.}}{2008}]%
        {vamplew2008limitations}
\bibfield{author}{\bibinfo{person}{Peter Vamplew}, \bibinfo{person}{John
  Yearwood}, \bibinfo{person}{Richard Dazeley}, {and} \bibinfo{person}{Adam
  Berry}.} \bibinfo{year}{2008}\natexlab{}.
\newblock \showarticletitle{On the limitations of scalarisation for
  multi-objective reinforcement learning of pareto fronts}. In
  \bibinfo{booktitle}{\emph{Australasian Joint Conference on Artificial
  Intelligence}}. Springer, \bibinfo{pages}{372--378}.
\newblock


\bibitem[\protect\citeauthoryear{Watkins}{Watkins}{1989}]%
        {watkins1989learning}
\bibfield{author}{\bibinfo{person}{Christopher John Cornish~Hellaby Watkins}.}
  \bibinfo{year}{1989}\natexlab{}.
\newblock \emph{\bibinfo{title}{Learning from Delayed Rewards}}.
\newblock \bibinfo{thesistype}{Ph.D. Dissertation}. \bibinfo{school}{King's
  College}, \bibinfo{address}{Cambridge, UK}.
\newblock


\bibitem[\protect\citeauthoryear{Wooldridge}{Wooldridge}{2009}]%
        {wooldridge2009introduction}
\bibfield{author}{\bibinfo{person}{Michael Wooldridge}.}
  \bibinfo{year}{2009}\natexlab{}.
\newblock \bibinfo{booktitle}{\emph{An introduction to multiagent systems}}.
\newblock \bibinfo{publisher}{John Wiley \& Sons}.
\newblock


\bibitem[\protect\citeauthoryear{Zhang, R\u{a}dulescu, Mannion, Roijers, and
  Now\'{e}}{Zhang et~al\mbox{.}}{2020}]%
        {Zhang2020OpponentAAMAS}
\bibfield{author}{\bibinfo{person}{Yijie Zhang}, \bibinfo{person}{Roxana
  R\u{a}dulescu}, \bibinfo{person}{Patrick Mannion},
  \bibinfo{person}{Diederik~Marijn Roijers}, {and} \bibinfo{person}{Ann
  Now\'{e}}.} \bibinfo{year}{2020}\natexlab{}.
\newblock \showarticletitle{Opponent Modelling for Reinforcement Learning in
  Multi-Objective Normal Form Games}. In \bibinfo{booktitle}{\emph{Proceedings
  of the 19th International Conference on Autonomous Agents and Multiagent
  Systems (AAMAS)}}.
\newblock


\end{thebibliography}


\end{document}